\documentclass[preprint,12pt]{elsarticle}
\usepackage{amssymb}
\usepackage[nodots]{numcompress}
\usepackage{subfigure}
\journal{Metamaterials}

\begin{document}

\begin{frontmatter}

\title{A comparative study of semiconductor-based plasmonic metamaterials}

\author[PU]{Gururaj V. Naik}
\ead{gnaik@purdue.edu}
\author[PU,DTU]{Alexandra Boltasseva\corref{cor}}
\ead{aeb@purdue.edu}
\cortext[cor]{Corresponding author, Phone: +1-765-4940301, Fax: +1-765-494-6951}
\address[PU]{Birck Nanotechnology Center and School of Electrical \& Computer Engineering,
Purdue University, IN 47907 USA}
\address[DTU]{DTU Fotonik, Technical University of Denmark, Lyngby 2800, Denmark}

\begin{abstract}
Recent metamaterial (MM) research faces several problems when using metal-based plasmonic components as building blocks for MMs. The use of conventional metals for MMs is limited by several factors: metals such as gold and silver have high losses in the visible and near-infrared (NIR) ranges and very large negative real permittivity values, and in addition, their optical properties cannot be tuned. These issues that put severe constraints on the device applications of MMs could be overcome if semiconductors are used as plasmonic materials instead of metals. Heavily doped, wide bandgap oxide semiconductors could exhibit both a small negative real permittivity and relatively small losses in the NIR. Heavily doped oxides of zinc and indium were already reported to be good, low loss alternatives to metals in the NIR range. Here, we consider these transparent conducting oxides (TCOs) as alternative plasmonic materials for many specific applications ranging from surface-plasmon-polariton waveguides to MMs with hyperbolic dispersion and epsilon-near-zero (ENZ) materials. We show that TCOs outperform conventional metals for ENZ and other MM-applications in the NIR.
\end{abstract}


\end{frontmatter}


\section{Introduction}
Plasmonics and the recent birth of metamaterials (MMs) \cite{negindex_veselago,perfectlens_pendry} (for recent reviews on optical metamaterials see, for example, \cite{reviewNP_shalaev,reviewSci_shalaev,book_wenshen}) and transformation optics (TO) \cite{reviewSci_shalaev,TO_pendry,schurig2006metamaterial} are currently driving the development of a family of novel devices with unprecedented functionalities ranging from subwavelength plasmonic waveguides and optical nanoresonators \cite{nature_barnes,review_maier_atwater,LSPR_review,plasmonicsOzbay} to superlenses \cite{perfectlens_pendry}, hyperlenses \cite{hyperlens_jacob,microscopy_engheta} and light concentrators \cite{TO_kildishev}. For plasmonic systems, metals have traditionally been the material of choice due to their ability to support collective oscillations of free electrons (called plasmons) and thereby, enabling the unique property of focusing light down to the nanoscale. However, as plasmonic devices operate at increasingly higher frequencies in the optical and telecommunications ranges, they begin to suffer from high losses arising in part from interband electronic transitions in the metals \cite{optprop_JC}. Even the metals with the highest conductivities (like silver and gold) suffer from large losses at optical frequencies \cite{optprop_JC}. These losses are detrimental to the performance of plasmonic devices, seriously limiting the feasibility of many plasmonic applications.\\

Currently, the best-known solution to the problem of losses is to use gain medium as a host material, thereby providing a means to compensate for the losses inherent to the plasmonic structures \cite{sudarkin1989excitation,nezhad2004gain,activeMM_noginov,gainSPP_berini}. It has been recently realized that energy can be transferred from a gain material to surface plasmons in metal nanostructures using stimulated emission. Unfortunately, even the largest gain provided by existing active materials is only barely enough to compensate losses in plasmonic materials (see for example \cite{NIM_klar}). However, if the plasmonic material losses could be reduced by a factor of two or three, the losses could be completely compensated by the gain medium.\\

Another limitation when using conventional metals for the plasmonic building blocks of MM devices is a negative real part of permittivity that is too large in the near-infrared (NIR) range and at telecommunication wavelength of 1.55 $\mu m$. Large negative values of real permittivity ($\epsilon'$) seriously limit the realization of TO devices because such devices often require similar magnitudes of $\epsilon'$ for their plasmonic and dielectric components \cite{TO_kildishev}. Dielectrics at optical frequencies have permittivity values on the order of 1, while for metals it is on the order of 10 or more. Thus, conventional metals are typically not good choices for many TO-based applications. Hence, devices such as hyperlens were realized only in the UV range \cite{hyperlens_zhang}.\\

Because of large losses and large magnitude of $\epsilon'$ inherent to the conventional plasmonic materials such as noble metals, it is clear that alternative plasmonic materials must be developed for practical applications of plasmonic and MM systems. Recently, we conducted a comprehensive study of various plasmonic materials including metals, metal alloys and heavily doped semiconductors evaluating the performance of each material for different plasmonic applications and outlining important aspects of their fabrication \cite{APM_LPR}. We showed that transparent conducting oxides (TCOs) such as aluminum zinc oxide (AZO), indium tin oxide (ITO) and gallium zinc oxide (GZO) are good candidates as plasmonic materials in the NIR range because they exhibit smaller losses and small negative $\epsilon'$ values in the NIR \cite{APM_LPR,AZO_RRL}. TCOs can be doped much higher than many other semiconductors (such as silicon) which gives them metal-like optical properties in the NIR. Plasmonic applications with TCOs were demonstrated with ITO in the NIR \cite{TCOspp_franzen,ITO_franzen}. However, there are no such reports on AZO. Hence, we briefly report the challenges involved in developing AZO as a plasmonic material in the NIR. AZO, GZO and ITO, together as TCOs, were recognized as good alternative plasmonic material candidates \cite{APM_LPR}. However, the design of specific plasmonic devices and the assessment of their performances are not reported. In this paper, we provide a comparative quantitative assessment of the performances of specific plasmonic MM-devices with TCOs as their plasmonic components rather than the conventional plasmonic metals. Our comparative study focuses on the following classes of devices: surface-plasmon-polariton (SPP) waveguides, localized surface plasmon resonance (LSPR) applications, hyperbolic metamaterials (HMM), TO devices and epsilon-near-zero (ENZ) applications.

\section{Development of Al:ZnO as a plasmonic material in the NIR}
 Zinc oxide and indium oxide have large solid-solubilities for dopants (Al, Ga in ZnO and Sn in $\mbox{In}_2\mbox{O}_3$) which enables high levels of doping \cite{SolidSolubility_yoon}. Large doping can result in high carrier concentration ($\approx 10^{21}cm^{-3}$) which is essential for achieving negative $\epsilon'$ in the NIR. Reaching such high carrier concentration is non-trivial due to several factors, such as the presence of defects in the material \cite{defects_AZO}. Zinc oxide and indium oxides are non-stoichiometric oxides having high defect densities at room temperature. The deposition conditions of these oxide semiconductors significantly influence the defect structures and thus impact their optical properties \cite{AZO_horwitz,AZO_hiramatsu,ITO_franzen}.\\

 Indium tin oxide (ITO) was demonstrated as a plasmonic material for wavelengths longer than 1.3 $\mu m$ \cite{TCOspp_franzen,switchingTCOs_atwater}. However, ITO has some drawbacks when compared to AZO. ITO typically forms amorphous films resulting in higher carrier scattering. AZO, on the other hand, forms polycrystalline films which exhibit higher carrier mobility and smaller optical losses than ITO \cite{review_gordon,ITO_AZO_comp}. Another drawback is that indium oxide is quite expensive due to limited natural reserves of indium. Due in part to these issues, recent research on TCOs has shifted the focus from ITO to AZO.\\

 \begin{figure}[htb]
\centering
\mbox{\subfigure{\includegraphics[width=6.66cm]{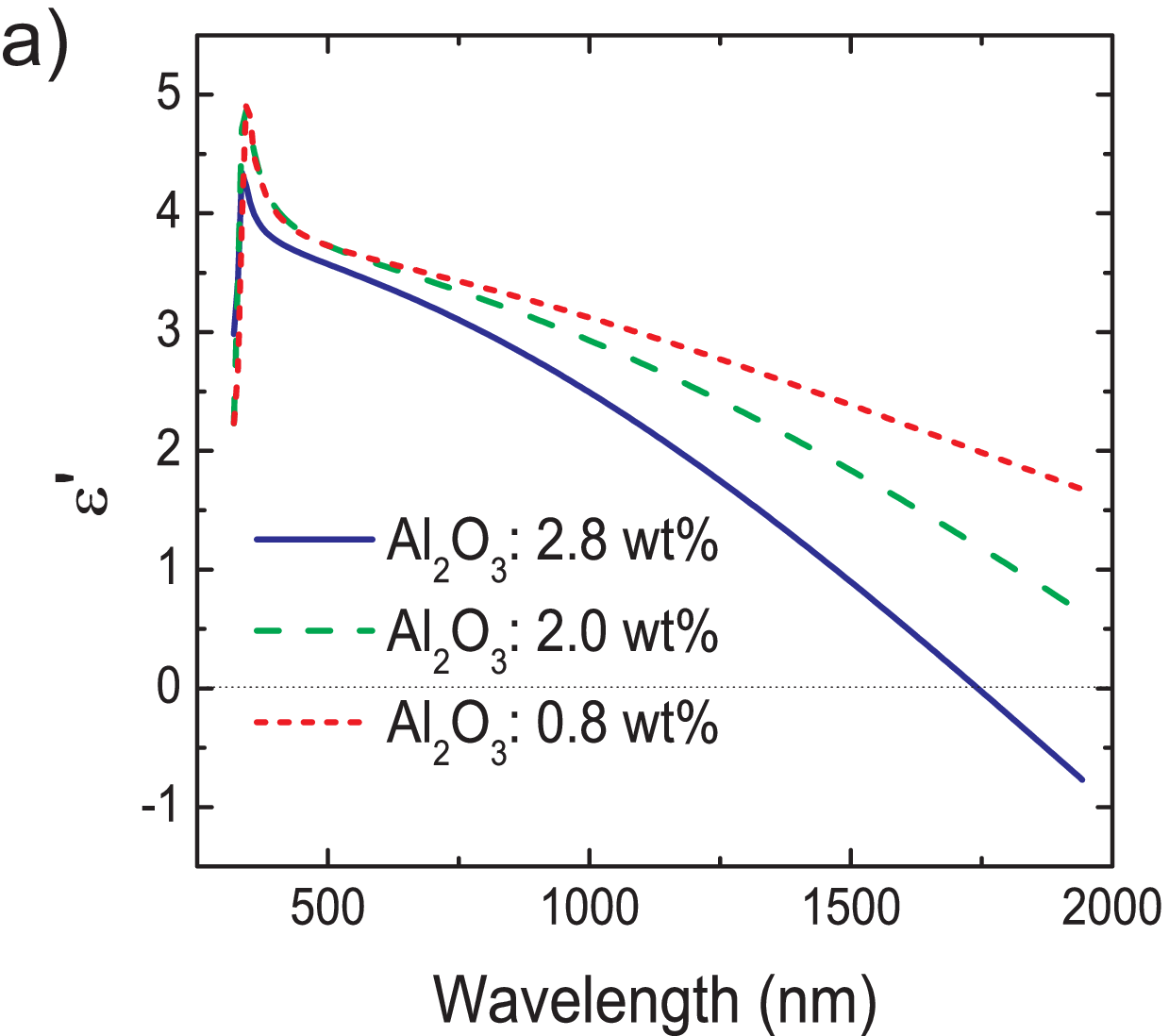}
\quad
\subfigure{\includegraphics[width=6.66cm]{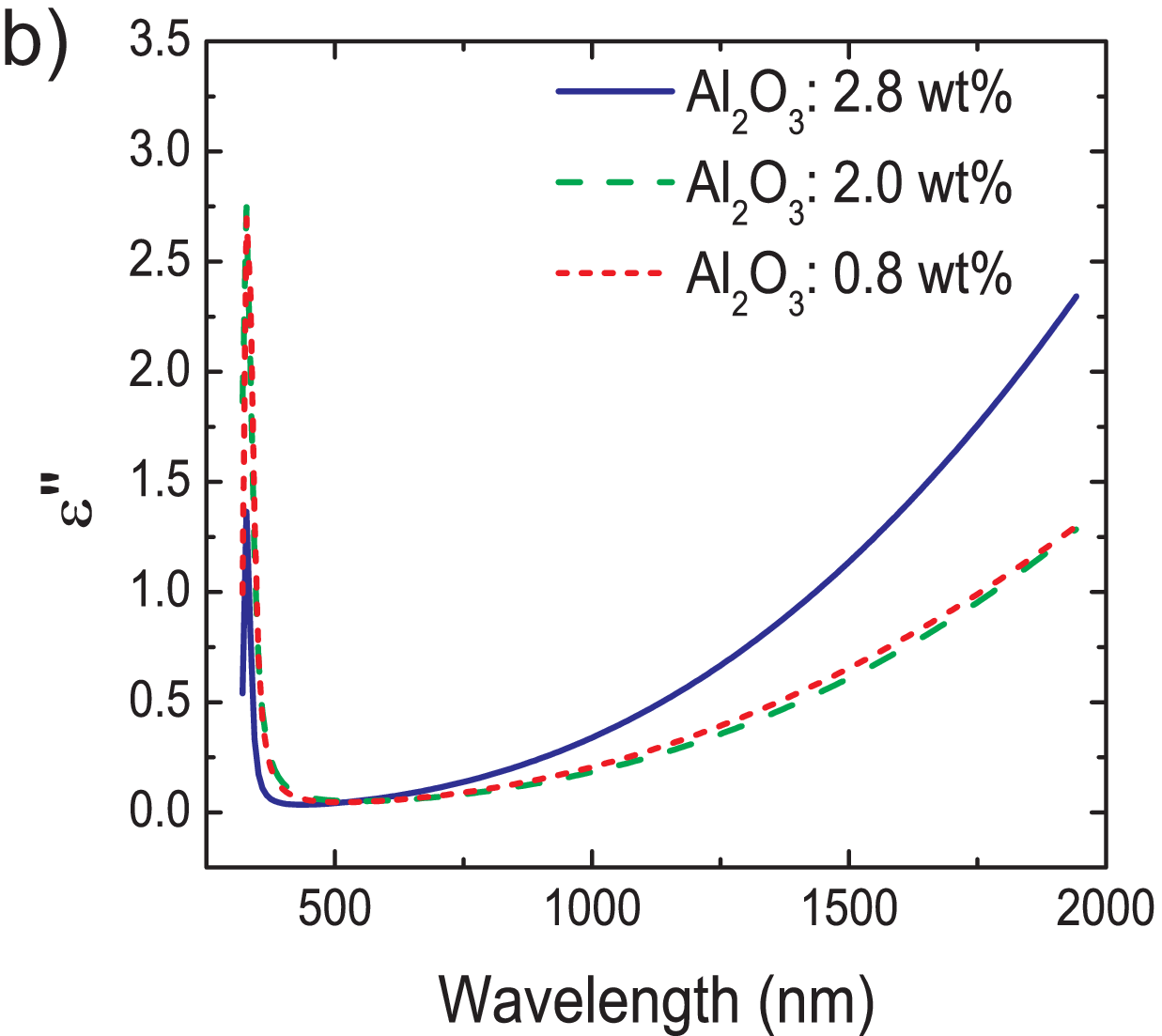} }}}
\caption{Dielectric function of pulsed laser deposited Al:ZnO films retrieved by ellipsometry measurements. The films were deposited with aluminum doping of 3.0 wt\% on c-sapphire substrates at deposition temperatures of 100 $^oC$ and 150 $^oC$. One of the films deposited at 150 $^oC$ was subjected to forming gas anneal at 300 $^oC$ for 2 hours.}
\label{fig1}
\end{figure}

High quality AZO films can be produced by sputtering or laser ablation (also called pulsed laser deposition (PLD)) deposition techniques \cite{reviewTCO_minami}. In our studies, we deposited AZO films by the PLD technique where high energy laser pulses from a KrF excimer laser ablates aluminum oxide and zinc oxide targets. The films are deposited at 100 $^oC$ and 150 $^oC$ in oxygen ambient on c-sapphire substrates. One of films was subjected to post-deposition anneal in 4:10 $H_2:N_2$ ambient at 300 $^oC$ for 2 hours. The films are optically characterized using a variable angle spectroscopic ellipsometer (J.A. Woollam Co.). Figures \ref{fig1}a and \ref{fig1}b show the dielectric function retrieved for AZO films with different processing conditions. The properties of the films are quite sensitive to the deposition and annealing conditions. Figures \ref{fig1}a and \ref{fig1}b show that lower deposition temperature decreases crossover wavelength (the wavelength at which $\epsilon'$ crosses zero) slightly but, increases damping losses. Annealing makes doping less efficient and renders AZO films less metallic in the NIR. In order to render AZO films metallic at 1.55 $\mu m$ wavelength, the cross-over wavelength should decrease further. This could be possible by increasing the doping efficiency which might require modifications to the deposition scheme such as ablating single target only. More details about the process optimization are provided in \cite{metaCongress_naik,spie_naik}.

\section{Comparative study}
 In this section, we quantitatively assess TCOs as alternate plasmonic materials in terms of their performance in various applications. Specifically, we compare AZO as a plasmonic material against conventional metals for NIR applications including SPP-waveguides, LSPR-applications, hyperbolic metamaterials, TO-devices and ENZ materials. For our comparison, we use the optical data for metals from Johnson and Christy \cite{optprop_JC}, and that for AZO is obtained from theoretical calculations based on data in Ref. \cite{AZO_hiramatsu,APM_LPR}.

\begin{figure}[htb]
\centering
\mbox{\subfigure{\includegraphics[width=6.66cm]{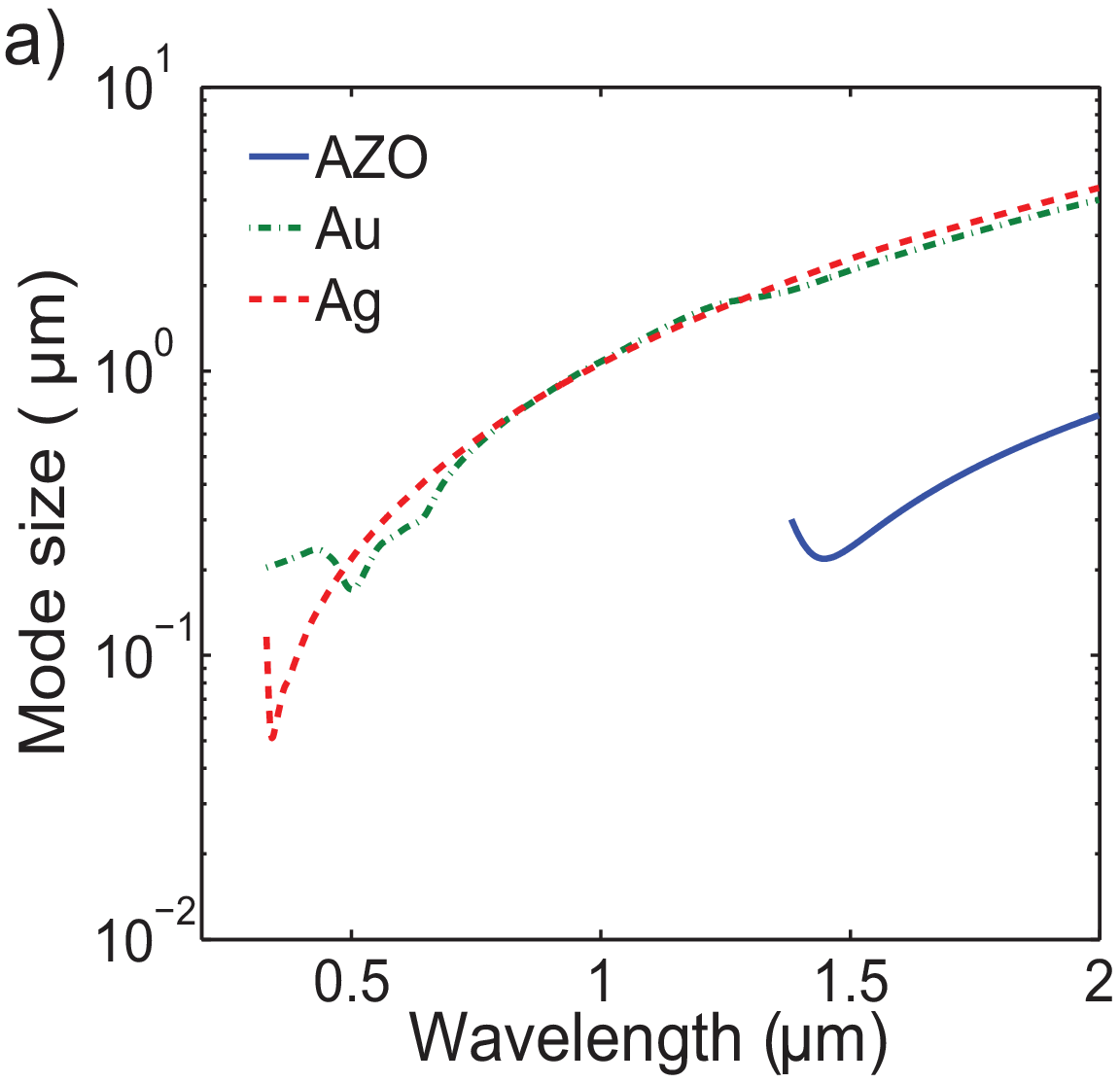}
\quad
\subfigure{\includegraphics[width=6.66cm]{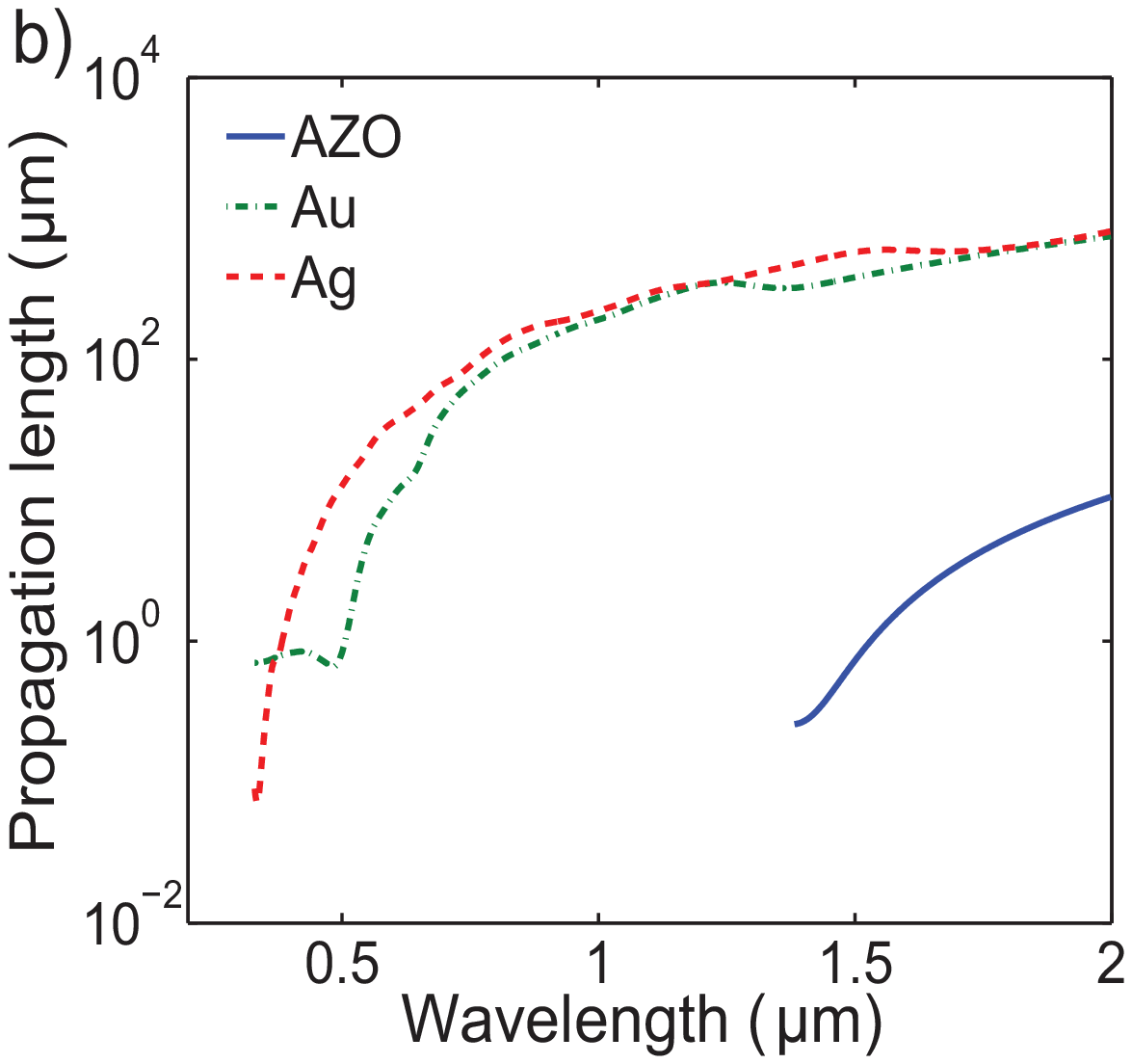} }}}
\caption{(a) Mode size and (b) propagation length of SPP on single interface: Au/air, Ag/air and TCO/air waveguide}
\label{fig2}
\end{figure}

\begin{figure}[h]
\begin{center}
\includegraphics[width=6.66cm]{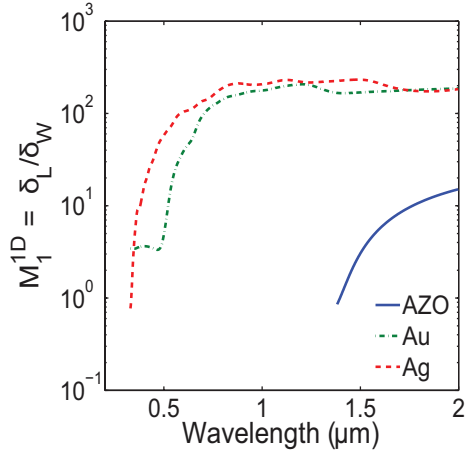}
\caption{Quality factor of single interface (Au/air, Ag/air, TCO/air) SPP-waveguide showing the trade-offs between propagation length and mode confinement}
\label{fig3}
\end{center}
\end{figure}

\begin{figure}[h]
\begin{center}
\includegraphics[width=6.66cm]{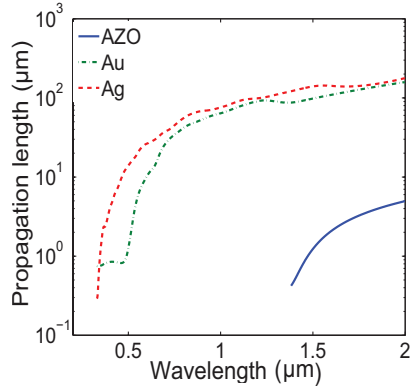}
\caption{Propagation length of the lowest order long-range SPP mode on metal-air-metal gap waveguide with gap size of 300 nm}
\label{fig4}
\end{center}
\end{figure}

\begin{figure}[h]
\begin{center}
\includegraphics[width=6.66cm]{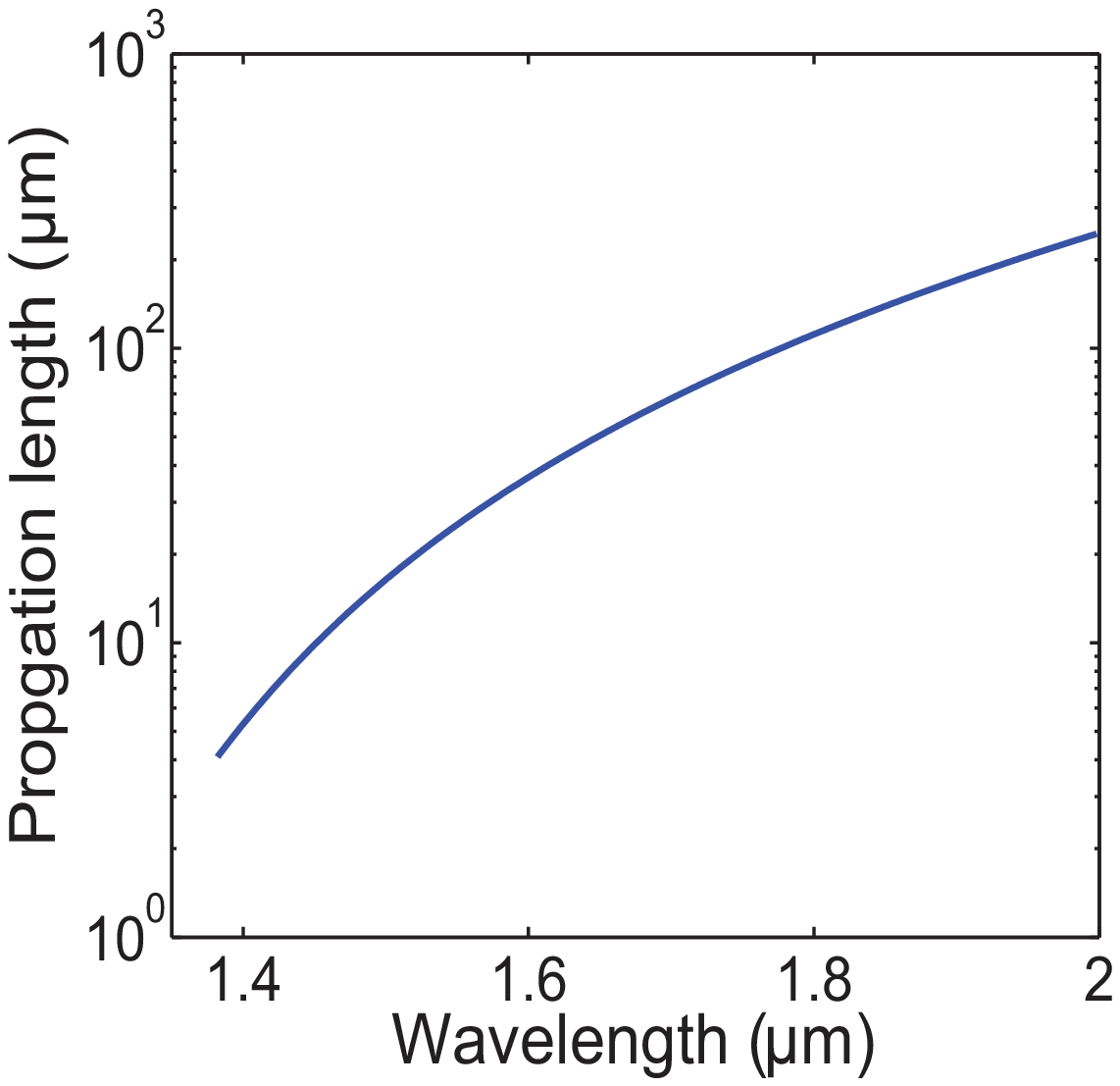}
\caption{Propagation length of the lowest order long-range SPP mode on air/AZO/air waveguide with AZO thickness of 100 nm}
\label{fig5}
\end{center}
\end{figure}

\subsection{SPP applications}

Carefully designed and fabricated TCOs could exhibit negative $\epsilon'$ in the NIR and hence, could be used for SPP waveguide applications. Consider a single interface of metal or TCO and air/vacuum. The SPP propagating on this interface has a 1/e propagation length given by $\delta_L=1/Im\{\beta_{SPP}\}$, where $\beta_{SPP}$ is the SPP-propagation constant. The mode size $\delta_W$ is given by Eq. \ref{eq_spp}, where $\delta_D$ and $\delta_m$ are the 1/e field decay lengths in the dielectric and metal, respectively, and $\epsilon_D$ and $\epsilon_m$ are the permittivity values of the dielectric and metal, respectively \cite{FOM_berini}.

 \begin{equation}
 \delta_W = \left\{ \begin{array}{ll}
 \delta_D & \mbox{for} |\epsilon_m|>e\epsilon_D\\
 \delta_D+\delta_D ln(e\epsilon_D/|\epsilon_m|) & \mbox{for} |\epsilon_m|<e\epsilon_D
 \end{array}
 \right.
 \end{equation}\label{eq_spp}

 Figures \ref{fig2}a and \ref{fig2}b show the mode size and propagation length of the single interface SPP waveguide for gold, silver and a typical TCO. Clearly, TCOs offer an order of magnitude better confinement than gold or silver for such SPP-waveguides in the NIR range. However, the propagation length for TCO/air SPP-waveguide is nearly two orders smaller than that for gold or silver waveguides. Thus, there exists a trade-off between propagation length and confinement. This trade-off is captured in Fig. \ref{fig3} where the quality factor of 1D waveguides, $M_1^{1D}=\delta_L/\delta_W$ \cite{FOM_berini} is plotted for TCO, gold and silver waveguides. The quality factor in the NIR is about an order of magnitude larger in the case of silver and gold waveguides than that for TCOs. This implies that for the same propagation length, metal waveguides could provide better confinement than an TCO waveguide. This fact can be appreciated better by considering a metal-air-metal gap waveguide (MIM waveguide \cite{slotWvgd_atwater}) with a 300 nm air gap. The mode size of this waveguide in the NIR is essentially 300 nm (the same as with an TCO/air single interface waveguide). However, the propagation lengths of the lowest order long-range mode (see Fig. \ref{fig4}) in silver and gold waveguides are about an order higher than that in single interface TCO/air waveguide and more than an order higher than in an TCO/air/TCO waveguide. Thus, TCOs are not good replacements for metals in the NIR for standard SPP applications.\\

Though TCOs are not better than noble metals for SPP waveguiding applications, they could have niche applications involving SPPs such as in SPP-based chemical sensing. TCOs are known to be good for gas sensing because they can oxidize many gaseous chemicals, thereby modulating the TCO-carrier concentration \cite{review_gas_sensing}. A change in the carrier concentration alters the dielectric function \cite{switchingTCOs_atwater}, which in turn sharply changes the surface plasmon resonance (SPR) conditions. Thus, highly sensitive SPR-based gas sensing could be an application where plasmonic TCOs could be employed. In such applications, propagation length of the SPP waveguide formed by TCOs could be improved by adopting an insulator-TCO-insulator geometry (IMI waveguide \cite{IMI_berini}). Figure \ref{fig5} shows the propagation length of the lowest order long-range SPP mode in air/AZO/air waveguide with a 100 nm AZO film. The propagation length is large enough to enable many niche applications such as SPP-based gas sensing.

\subsection{LSPR applications}
Localized SP resonances in metallic nanospheres are used in many applications, such as sensing \cite{LSPR_review}. Silver and gold nanoparticles were demonstrated to have LSPR modes at blue and green wavelengths, respectively \cite{AgAu_LSPR}. However, the resonances of these spherical metal nanoparticles are limited only to the visible spectrum. For longer wavelengths in the NIR, nanospheres of TCO nanoparticles could be efficient \cite{ITO_LSPR}. The quality factor of the resonance of spherical nanoparticles is $-\epsilon_m'/\epsilon_m"$ \cite{APM_LPR}. Using this expression, the quality factor of silver nanospheres turns out to be about 6 at a wavelength of 400 nm. Similarly, gold nanospheres have LSPR quality factors of about 4.5 at wavelengths near 500 nm, and TCO nanoparticles have a quality factor of about 2.4 at a wavelength of 2 $\mu m$. Though TCOs have slightly lower quality factors, they enable LSPR applications in the NIR. Adopting a nanoshell structure for metals (Au or Ag) also enables LSPR applications in the NIR with larger quality factors \cite{LSPR_shells}. However, TCO nanospheres might offer easier fabrication of plasmonic structures than metal nanoshells for LSPR applications in the NIR.

\subsection{Hyperbolic metamaterials}
\begin{figure}[h]
\begin{center}
\includegraphics[width=6.66cm]{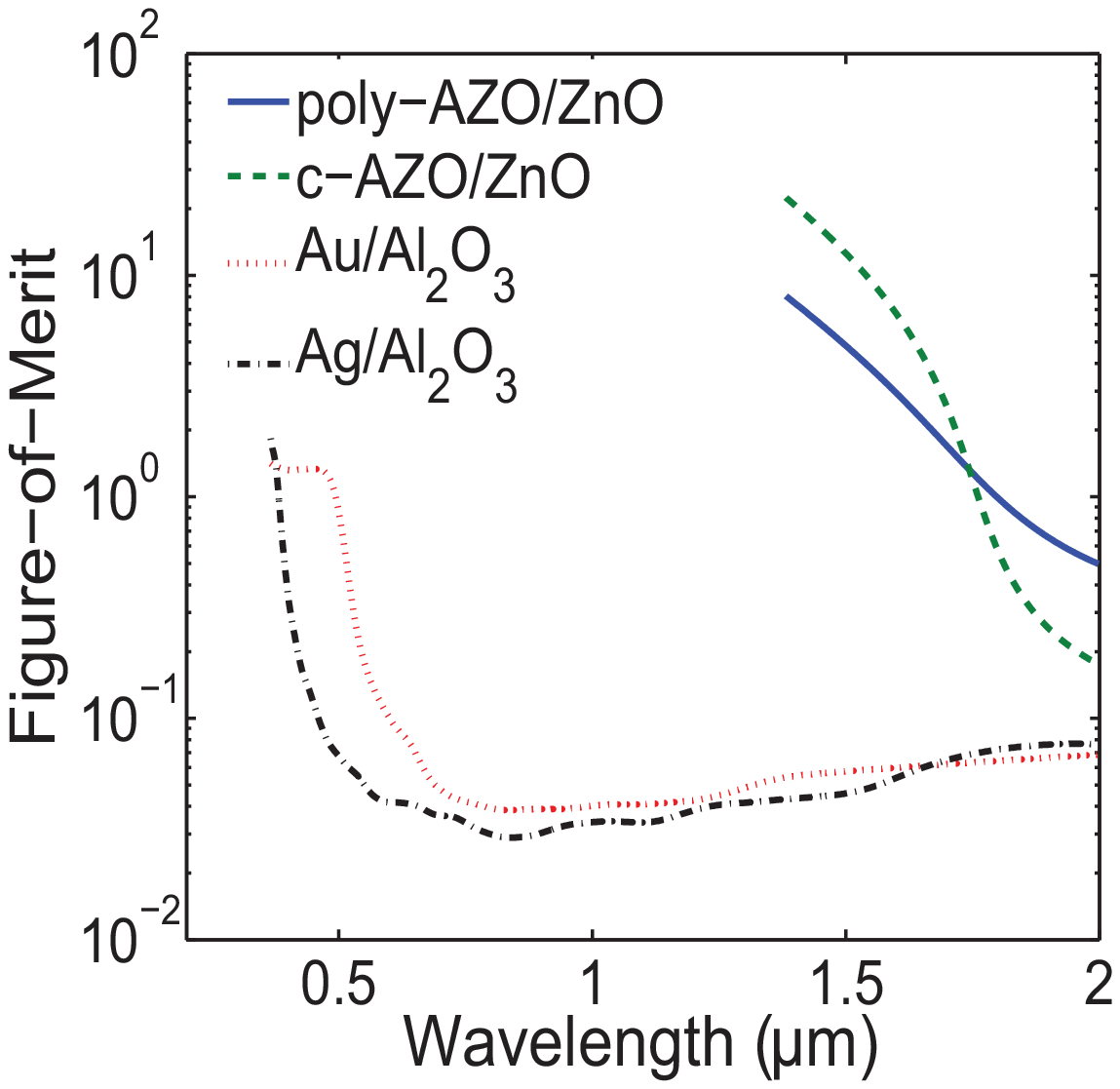}
\caption{Figure-of-merit of hyperbolic metamaterial made from sub-wavelength, alternating layers of metal and dielectric materials: polycrystalline AZO/ZnO, single crystal AZO/ZnO, Au/$\mbox{Al}_2\mbox{O}_3$ and Ag/$\mbox{Al}_2\mbox{O}_3$}
\label{fig6}
\end{center}
\end{figure}

Hyperbolic metamaterials (HMMs) are receiving great attention currently due to their special properties. HMMs can be used as hyperlens devices for sub-diffraction imaging since they support the propagation of extremely high-\emph{k} waves \cite{hyperlens_jacob,hyperlens_zhang}. Recently, HMMs were shown to exhibit a broadband singularity in the photonic density of states \cite{BBpurcell_jacob}. This phenomenon is very useful in applications such as single-photon guns because HMMs can enormously increase the spontaneous emission rate of fluorophores \cite{PDOSengg_jacob}. All of these properties arise from the hyperbolic dispersion exhibited by such materials.

One of the easiest ways of achieving hyperbolic dispersion is to stack many sub-wavelength, alternating layers of dielectric and metal \cite{hyperlens_jacob}. In the effective medium approximation, such layered stack produces in-plane permittivity ($\epsilon_{\parallel}$) and out-of-plane permittivity ($\epsilon_{\perp}$) of different signs. This is the basis of producing hyperbolic dispersion. In reality, such a hyperbolic metamaterial would suffer from large losses due to the losses in the constituent metal. Minimizing the losses by optimizing the choice of materials and the design of the HMM are crucial steps in achieving reasonable performance of HMM-based devices. The performance of HMM devices can be quantified based on the figure-of-merit (FOM) as suggested in Ref. \cite{negref_hoffman}: $FOM=Re\{k_{\perp}\}/Im\{k_{\perp}\}$. Adopting this definition of the FOM for HMMs, in Fig. \ref{fig6} we show the HMM figure-of-merit for four different metal/dielectric layered systems: polycrystalline AZO/ZnO, single-crystal AZO/ZnO, Ag/$\mbox{Al}_2\mbox{O}_3$ (used in \cite{hyperlens_zhang}) and Au/$\mbox{Al}_2\mbox{O}_3$. Clearly, the system based on zinc oxide significantly outperforms the metal-based system in the NIR. Note that, in reference to \cite{AZO_RRL}, the FOM of any metal/dielectric layered system in the NIR is almost zero. Therefore, based on the FOM, TCO-based HMMs are more attractive in the NIR than metal-based HMM designs. Thus, TCOs are valuable in enabling HMM applications in the NIR and the ever-important telecommunication wavelength.

\subsection{Transformation Optics and ENZ applications}
As mentioned before, for a practical and efficient design of TO-devices, the real permittivity of the metallic components must be of the same order as that of dielectric components \cite{TO_pendry, TO_kildishev}. Since none of the traditional metals satisfy this condition in the NIR, these metals are not good plasmonic candidates for TO applications. However, TCOs are ideally suited for TO applications because their losses are smaller than conventional plasmonic metals and their real permittivity values are negative and small in the NIR.

\begin{figure}[h]
\begin{center}
\includegraphics[width=6.66cm]{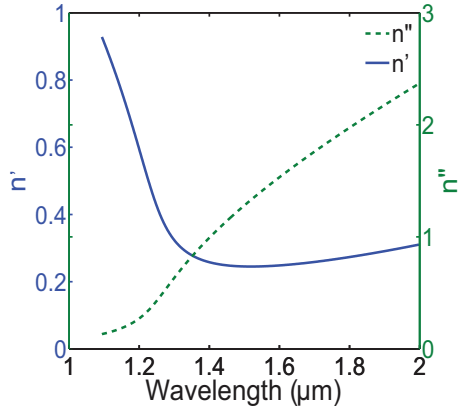}
\caption{Refractive index ($n=n\mbox{'}+in\mbox{"}$) of AZO in the NIR: n' reaches as low as 0.2 where ENZ applications could be possible.}
\label{fig7}
\end{center}
\end{figure}

The realization of ENZ devices requires materials with both low loss and small real permittivity. ENZ applications were originally proposed by Engheta \emph{et al.} \cite{enz_engheta} in 2006 and were later developed to include many novel functionalities such as providing optical isolation between nano-optical circuit elements \cite{nanocircuit_engheta}, enhancing the directivity of nano-radiators \cite{extraordinaryT_engheta} and producing bulk impedance-matched materials \cite{radpattern_engheta}. Many dielectrics that have $\epsilon$-crossover points in the THz were identified as potential materials \cite{radpattern_engheta}. However, TCOs have $\epsilon$-crossover points in the NIR, meaning these materials could behave as ENZ materials in this range. The refractive index is a good parameter to indicate the effectiveness of these materials for ENZ applications. Figure \ref{fig7} shows the refractive index ($n=n\mbox{'}+in\mbox{"}$) of AZO in the NIR. The real part of the index goes as low as 0.2, while the imaginary part is about 1.5. Though these losses are not yet low enough, the real index is quite small and could enable some ENZ applications. Epsilon-zero materials are known to transmit only normally incident waves \cite{radpattern_engheta}. However, ENZ materials have a non-zero acceptance angle. With $n\mbox{'}$ =  0.2, a $\lambda/10$ slab of AZO would have an acceptance angle ($\Delta \theta/2$, see Eq. 6 of Ref. \cite{radpattern_engheta}) of about $18^o$. This suggests that AZO as an ENZ-shield would be a leaky insulator between two nanophotonic circuit elements but, with a small leakage \cite{nanocircuit_engheta}.

\section{Conclusions}
With the rapid development of the metamaterial field, it is clear that there is not a single plasmonic material that can be used as a suitable building block for all applications and at all frequencies. Currently, the applications of metamaterial-based devices are limited to the low-loss regions of available metals, which are far away from the important telecommunication wavelength. With the open-ended problem of losses in many plasmonic and metamaterial devices that are based on silver and gold, we have proposed alternative plasmonic materials based on conductive oxides that can be used for a new class of optical metamaterials. Aluminum zinc oxide is a low-loss, plasmonic material in the NIR with losses that are nearly five times smaller than those in silver. In this work, we assessed aluminum zinc oxide as an alternative plasmonic material in terms of its suitability for device applications in the NIR. Even though conventional metal outperforms the conducting oxides for SPP-waveguiding and LSPR applications, transparent conducting oxides can be invaluable in the realization of hyperbolic metamaterials, transformation optics and epsilon-near-zero devices in the NIR wavelength range. Thus, transparent conducting oxides hold the promise of real-life metamaterial applications and helping to unravel the interesting physical phenomena in a new generation of metamaterial and transformation-optics devices \cite{additional}.

\section*{Acknowledgments}
We thank Prof. Vladimir M. Shalaev for the useful discussions.
This work was supported in part by ONR MURI on ``Large-Area 3D Optical Metamaterials with Tunability and Low Loss.''



\end{document}